\documentclass[10pt,letterpaper,twocolumn]{article}
\usepackage[english]{babel}
\usepackage{graphicx}

\newcommand{\alt}{\mathbin{\lower 3pt\hbox
   {$\rlap{\raise 5pt\hbox{$\char'074$}}\mathchar"7218$}}}
\newcommand{\agt}{\mathbin{\lower 3pt\hbox
   {$\rlap{\raise 5pt\hbox{$\char'076$}}\mathchar"7218$}}}

\begin{document}

\setcounter{footnote}{0}
\setcounter{equation}{0}
\setcounter{figure}{0}
\setcounter{table}{0}

\title{\large\bf Spectral analysis of universal
conductance fluctuations }

\author{\small  I. M. Suslov \\
\small P.L.Kapitza Institute for Physical Problems,
119334 Moscow, Russia \\
\small E-mail: suslov@kapitza.ras.ru\\
{}\\
\parbox{150mm}{\footnotesize \,  Universal conductance
fluctuations  are usually observed in the form of aperiodic
oscillations in the magnetoresistance of thin wires as a function
 of the magnetic field $B$. If such oscillations are completely
 random at scales exceeding $\xi_B$, their Fourier analysis
 should reveal a white noise spectrum at frequencies below
 $\xi_B^{-1}$.
 Comparison  with the results for 1D systems suggests another scenario: according to it,
 such
 oscillations are due to the superposition of incommensurate
 harmonics and their spectrum should contain discrete
 frequencies. An accurate Fourier analysis of the classical
 experiment by Washburn and Webb reveals a practically discrete
 spectrum in agreement with the latter scenario.  However, this
 spectrum is close in shape to the discrete white noise spectrum
 whose properties are similar to a continuous one.
More detailed analysis reveals the existence of the continuous
component, whose smallness is explained theoretically.
A lot of qualitative results are obtained, which confirm
the presented picture. The distribution of phases,
frequency differences and the growth exponents agree with
theoretical predictions. Discrete frequencies depends
weakly on the treatment procedure. The discovered shift
oscillations confirm the analogy with 1D systems.
Microscopical estimates show agreement of
the obtained results
with geometrical dimensions of the sample.
} }

\date{}
\maketitle

\textwidth 6.4 in
\textheight 8.5 in

\setcounter{footnote}{0}
\setcounter{equation}{0}
\setcounter{figure}{0}
\setcounter{table}{0}

\begin{center}
{\bf 1. Introduction}
\end{center}

Universal conductance fluctuations
[1--4] are usually observed in the form of aperiodic
 oscillations in the magnetoresistance of thin wires as a
 function of the magnetic field $B$  \cite{5} (Fig.\,1)
(see \cite{200,201} for review).
The fluctuation picture looks random, but is completely
reproducible
from one measurement to another.
It is related with a specific realization of the random
potential and changes completely, if the sample is heated till
sufficiently large temperature, at which the impurities become
movable and a new impurity configuration
arises ("magnetic fingerprints").

 According to the theory [1--4], the
 conductance $G(B)$ at a given magnetic field $B$ undergoes
 fluctuations of the order of
 $e^2/h$ under the variation of the impurity
 configuration; fluctuations in $G(B)$ and $G(B+\Delta B)$ are
 statistically independent, if $\Delta B$ exceeds a certain
 characteristic scale $\xi_B$.  It is reasonable to expect that
 oscillations in $G(B)$ are completely random at
 scales exceeding $\xi_B$. Then, their Fourier analysis should
 reveal a white noise spectrum (i.e., frequency-independent
 plateau) at frequencies below $\xi_B^{-1}$.

 Comparison with the results for 1D systems suggests
 another scenario \cite{6}.  A magnetic field perpendicular to a
thin wire creates a quadratic potential along this wire \cite{7},
 which effectively restricts the length of the system $L$; hence,
 the variation of the magnetic field is similar to the variation
 of $L$. The resistance $\rho$ of a one-dimensional system is a
 strongly fluctuating quantity and the form of its distribution
 function $P(\rho)$ essentially depends on the first several
 moments.
\begin{figure}[t]
\centerline{\includegraphics[width=3.0 in]{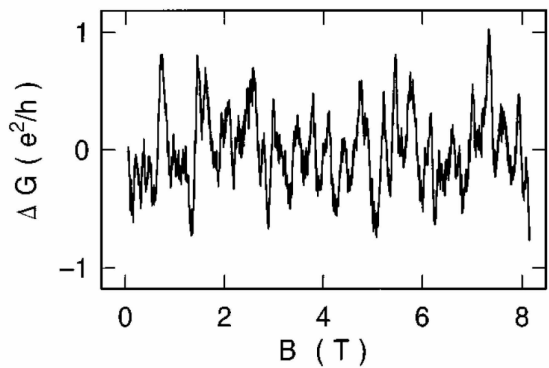}}
\caption{\small Conductance of the thin Au wire against the
magnetic field  \cite{5}.  }
\label{fig1}
\end{figure}
Indeed, the Fourier transform of $P(\rho)$ specifies
the characteristic  function
$$
{F}(t)=\left\langle e^{i\rho t} \right\rangle
=\sum_{n=0}^{\infty} \frac{(it)^n}{n!}\left\langle \rho^n
\right\rangle \,,
\eqno(1)
$$
 which is the generating function of the moments $\left\langle
 \rho^n \right\rangle$. If all moments of the distribution are
 known, the function ${F}(t)$ can be constructed using them, and
 the function $P(\rho)$ is then determined by the inverse Fourier
 transform. If an increase in the moments $\left\langle \rho^n
 \right\rangle$ with $n$ is not too fast, the contributions of
 higher moments are suppressed by a factor of $1/n!$, whereas
 first several moments are significant. These moments are
 oscillating functions of $L$,
$$
\left\langle \rho \right\rangle=a_1(L)+b_1(L)\cos(\omega_1
L+\varphi_1),
\eqno(2)
$$
$$
\left\langle \rho^2 \right\rangle=a_2(L)+b_2(L)\cos(\omega_2
L+\varphi_2)+
$$
$$+
b_3(L)\cos(\omega_3
L+\varphi_3)\,, \quad\mbox{etc.,}
$$
where $a_s(L)$ and $b_s(L)$ are monotonic functions. The
reason is that the growth exponent for $\left\langle \rho^n
\right\rangle$ is determined by the $(2n\!+\!1)$th order
 algebraic equation \cite{6}, one of whose
 root is always real, whereas
 the other roots are complex
for energies in the allowed band.
 Consequently, there are $n$ pairs of complex conjugate
 roots. An expression for $\langle
 \rho^n \rangle$ contains a linear combination of the
 corresponding exponents, and complex roots provide
the existence of $n$ oscillating terms.
The frequencies $\omega_s$ are usually incommensurate,
 but their incommensurability vanishes in the deep of
 the allowed band at weak disorder (Sec.\,9).
 According to this picture, oscillations in
 $G(B)$ shown in Fig.\,1 are determined by the superposition of
 incommensurate harmonics and their Fourier spectrum should
 contain discrete frequencies. This picture is indirectly
 confirmed by the experimental data obtained in \cite{8} and
 cited in \cite{6}, according to which the distribution function
 $P(\rho)$ is not stationary, but demonstrates systematic
 aperiodic variations.

It is clear from the above that the Fourier analysis of the
function $G(B)$ makes it possible to establish which of two
scenarios is more adequate.
It will be shown below that such analysis results in the
spectrum, which looks purely discrete (Sec.\,2), indicating
validity of the second conception.
Nevertheless, no
contradictions arise with the
diagrammatic results [1--4], since the form of the
spectrum in whole is close to the discrete white noise,
whose properties are close to the continuous one. More
detailed analysis (Sec.\,4) reveals the existence of the
continuous component, whose smallness is explained
theoretically in Sec.\,5. Dependence of results of the treatment
procedure is discussed in Sec.\,3. In spite of the
evident problems, arising due to deviations from the optimal
regime, the discrete frequencies of the spectral lines
manifest the miraculous stability, proving their objective
origin. Analysis of the real and imaginary parts of the Fourier
transform $F(\omega)$ of the function $G(B)$ (Fig.\,1) reveals
the existence of quick oscillations, related with the shift of
$B$ from its "natural" origin, whose nature is discussed in
Sec.\,6. After exclusion of quick oscillations we study the
distribution of phase shifts for the discrete harmonics, which
does not contradict their expected
stochastization (Sec.\,7).  Positions
of extrema for ${\rm Re}F(\omega)$ and ${\rm Im}F(\omega)$
differ from those for $|F(\omega)|$ indicating the existence
of the exponential growth of harmonics expected from the analogy
with 1D systems (Sec.\,8). The distribution of the growth exponents
and frequency differences corresponds to the theoretical
expectations for the metallic regime  (Sec.\,9).
Microscopical estimates (Sec.10) indicate an agreement of the
presented picture
with geometrical dimensions of the sample.
A brief communication on the obtained results was published
previously \cite{11a}.

\begin{center}
{\bf 2. Fourier spectrum of aperiodic oscillations}
\end{center}

A Fourier analysis of the function $G(B)$  (Fig.\,1)
cannot be carried out directly, since the abrupt
cutoff of experimental data gives rise to
slowly decaying oscillations  in its spectrum
 and chaotization of
the latter\,\footnote{\,Figure 14 in \cite{5} shows the Fourier
 spectrum of a thin wire in comparison with the spectrum of a
 small ring; the latter contains additional oscillations caused
 by the Aharonov--Bohm effect. However, aperiodic oscillations
were not discussed in this place
 and their spectrum, which is
 chaotic because of the sharp cutoff, was roughly approximated by
 the authors in the form of the envelope of oscillations. This is
 obvious from comparison with Figs. 12 and 13 in \cite{5}, where
 chaotic oscillations are clearly seen.}. To obtain clear
 results, it is necessary to use a proper smoothing
 function.
 Let discuss a situation in more details.

Let the function $f(x)$ be the superposition of discrete
harmonics and be real. Then,
$$
f(x)=\sum_s A_s {\rm e}^{i\omega_s x}=
\frac{1}{2}\sum_s \left[ A_s {\rm e}^{i\omega_s x}
+A_s^* {\rm e}^{-i\omega_s x} \right] \,,
\eqno(3)
$$
 where the frequencies $\omega_s$ can be considered as positive
 without loss of generality. Then, the Fourier transform
of $f(x)$  has the form
$$
F(\omega)=\pi \sum_s \left[ A_s \delta(\omega+\omega_s)
+A_s^* \delta(\omega-\omega_s) \right]  \,,
\eqno(4)
$$
and its modulus
$$
|F(\omega)|=\pi \sum_s |A_s|\left[
\delta(\omega+\omega_s) +\delta(\omega-\omega_s) \right]
\eqno(5)
$$
\begin{figure*}[t]
\centerline{\includegraphics[width=5.5 in]{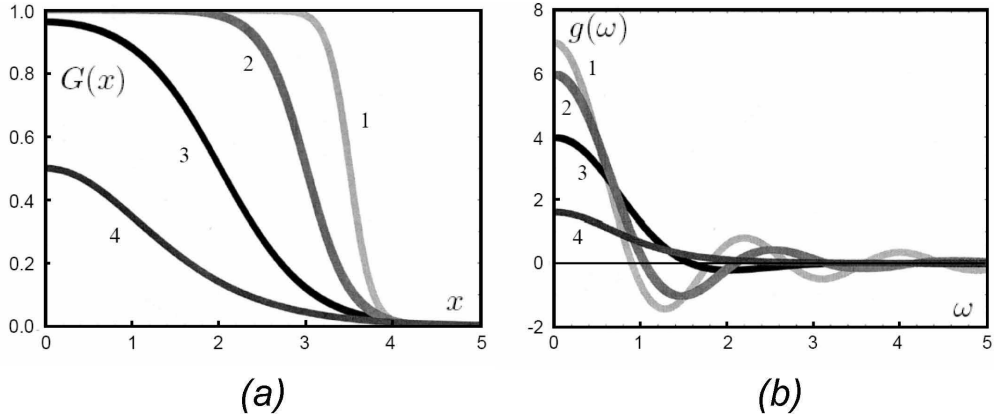}}
\caption{\small (a)  Function $G(x)$ given by Eq.\,9, and (b)
its Fourier transform $g(\omega)$  at ({\it 1}) $\mu=3.5$,
$T=0.125$; ({\it 2}) $\mu=3$, $T=0.25$; ({\it 3}) $\mu=2$,
$T=0.5$; and ({\it 4}) $\mu=T\ln 2$, $T=0.8$.
} \label{fig2}
\end{figure*}
 depends only on the intensities of spectral lines and does not
 contain information on phase shifts in the corresponding
 harmonics.  Since $|F(\omega)|$ is an even function, it is
 possible to consider only positive $\omega$ values and to omit
 the first delta function in Eq.\,5.

 Since the function $f(x)$ can be experimentally measured only in
 a
 finite $x$ range, we in practice have
$$
f(x)=\frac{1}{2}\sum_s \left[ A_s {\rm e}^{i\omega_s x}
+A_s^* {\rm e}^{-i\omega_s x} \right] G(x)\,,
\eqno(6)
$$
 where the function $G(x)$ is unity within the working
 range
 and zero beyond it; further, it will be smoothed. Then, instead
 of Eq.\,4, we obtain
$$
F(\omega)=\frac{1}{2} \sum_s \left[ A_s g(\omega+\omega_s)
+A_s^* g(\omega-\omega_s) \right]   \,,
\eqno(7)
$$
 where $g(\omega)$ is the Fourier transform of
 $G(x)$, which  is real for even function $G(x)$.
   Thus, the restriction of the working
 range
 leads to the replacement of delta functions by
 spectral lines with finite widths. If discrete frequencies are
 well separated and the function $g(\omega)$ is strongly
localized
 near zero, one can neglect the overlapping of
 functions $g(\omega\pm\omega_k)$ and write at positive
 frequencies
$$
|F(\omega)|^2 \approx \frac{1}{4} \sum_s  |A_s|^2
g^2(\omega-\omega_s)  \,.
\eqno(8)
$$
The function $|F(\omega)|^2$ (so-called  power spectral density
\cite{9}) more adequately characte- rizes the relative
 contribution of different harmonics,
since  the integral of
 this function over all frequencies is equal to the integral of
 $|f(x)|^2$ over all $x$ values.  Consequently, change in  the
 spectrum of $f(x)$ at fixed rms fluctuations results in the
 redistribution of intensities between different frequencies at
 the conservation of the total spectral power.

 It is easy to see that to obtain a clear picture in the case of
 a discrete spectrum, it is necessary to have a possibly narrower
 shape of spectral lines
 determined by $g(\omega)$,
  which can be achieved
 by the appropriate choice of the function $G(x)$. The general
 strategy is determined by the properties of integrals of rapidly
 oscillating functions \cite{10}. If
 the function $f(x)$ has discontinuity,
 its Fourier transform decreases at high frequencies as
 $1/\omega$; if the $n$th derivative is discontinuous,
 then $F(\omega)\sim \omega^{-n-1}$.  The Fourier transform of a
 smooth function $f(x)$  is calculated by shifting the contour of
 integration to a complex plane and is determined by the nearest
 singularity or saddle point, which leads to the dependence
 $F(\omega)\sim \exp(-\alpha \omega)$. If the regular function is
 obtained by means of a weak smoothing of
 a singularity, the $\alpha$ value is small and
 the exponential is manifested
 only at very high frequencies, whereas the behavior
 corresponding to the singular function holds in the remaining
 region.  In our case, it is necessary to smooth the
 discontinuity  of $G(x)$.  It should be clear that weak
 smoothing is  inefficient, while strong smoothing leads to small
 values of $G(x)$ near the boundaries of the working range and to
 loss of experimental information; so, a reasonable compromise is
  required.

Let  $G(x)$ be the $x$-symmetrized Fermi
function
$$
G(x)=\frac{1}{1+{\rm e}^{(x-\mu)/T}+{\rm e}^{(-x-\mu)/T}}
=
$$
$$=
\frac{1}{1+2{\rm e}^{-\mu/T}{\rm cosh}(x/T)} \,,
\eqno(9)
$$
whose
Fourier transform is given by the integral
$$
g(\omega)=\int\limits_{-\infty}^{\infty}\frac{{\rm e}^{i\omega x}\,dx}
{b\,{\rm cosh}\beta x +c}= \frac{2\pi}{b\beta{\rm sinh} x_0}
\frac{\sin(\omega x_0/\beta)}{{\rm sinh} (\omega\pi/\beta)}\,,
$$
$$
\qquad x_0={\rm arccosh}(c/b)  \,.
\eqno(10)
$$
If $x=B-\mu_0$ is chosen in our case, experimental data
correspond to the interval
$|x|\le \mu_0$ with $\mu_0=4$ (in units
of tesla).
As a rule we accepted $\mu=\mu_0-4T$, which ensures the small
value $G(\mu_0)\approx 0.02$ at
boundaries of the interval.
As clear from Fig.\,2, the behavior
$g(\omega)=2\sin{\mu\omega}/\omega$ characteristic of the sharp
cutoff prevails at small $T$  values, when smoothihg is weak
(lines {1} and {2}).  The extremal smoothing corresponds to
$\mu=T\ln{2}$,  $x_0=0$, when
$$
g(\omega)=\frac{2\pi T^2\omega}{ \sinh{\pi T\omega} }
\eqno(11)
$$
\begin{figure}[h]
\centerline{\includegraphics[width=3.8 in]{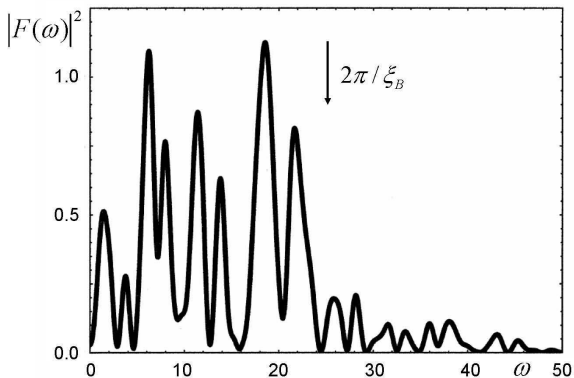}}
\caption{\small Fourier analysis of the experimental data
shown in Fig.\,1 for the smoothing function (9) with $\mu=T\ln
2$, $T=0.8$. Here values of $F(\omega)$  are multiplied by 10, as
well as in subsequent figures.
} \label{fig3}
\end{figure}
and oscillations disappear completely (line {4}).
It seems reasonable to choose $\mu=2$ and
$T=0.5$ (line {3}), as was made in the paper \cite{11a}; in this
case, about 50\% of experimental data are effectively used, while
the line shape is approximately the same as in the case of the
extremal smoothing.
In the present paper the choice (11) is more convenient;
it uses slightly less information but the form of the spectrum
is practically the same as in the paper \cite{11a}.

The spectral analysis of experimental data (Fig.\,1) was produced
by calculation of the Fourier integral in the region $|x|<\mu_0$
with the indicated smoothing function. The corresponding results
are shown in Fig.\,3. The spectrum
 obviously consists of discrete lines, which confirms the
 second scenario given in beginning.
 However, the spectrum in the
 range $\omega\alt 2\pi/\xi_B$ (where $\xi_B$ was estimated as
 the average distance between neighboring maxima or minima in
 Fig.\,1)\,\footnote{\,Under processing, Fig.\,1 was strongly
 magnified and digitized by hand.
It was revealed
 that sharp spikes in Fig.\,1 are due to vertical dashes
 indicating uncertainty of the data,
 whereas the   experimental dependence is in fact smooth.} is
 similar to discrete white noise: in a rough approximation, the
 lines are equidistant and their intensities are more or less
 the same.  Since the sum over frequencies is often approximated
 by an integral, discrete white noise does not differ in many
 properties from continuous white noise.
Let, for example,
$$
F(\omega)=\pi \sum_s \left[ A_s \delta(\omega+\omega_s)
+A_s^* \delta(\omega-\omega_s) \right] H(\omega) \,,
\eqno(12)
$$
 where the frequencies $\omega_s$ are equidistant
 ($\omega_s=s \Delta$), the amplitudes $A_s$ are the
 same in modulus
 ($|A_s|=A$) and have completely random phases, while
 $H(\omega)$ is an even function restricting the
 spectrum to the range $|\omega|\alt \Omega$.  Then, determining
 $f(x)$ by means of the inverse Fourier transform, we obtain the
 correlation function
$$
\langle f(x)f(x')\rangle=\frac{1}{2}\sum_s A^2 H^2(\omega_s)
{\rm e}^{i\omega_s (x-x')} \approx
$$
$$
\approx
\frac{1}{2} A^2 \Delta^{-1} h(x-x')    \,,
 \eqno(13)
$$
  where $h(x)$ is the Fourier transform of  $H^2(\omega)$.
 If the function $H(\omega)$  is smooth, $h(x)$
 decreases exponentially at a scale of $\Omega^{-1}$ in agreement
 with the diagrammatic results [1--4].

One can see that the obtained results  reconcile
 two  alternative scenarios described at the beginning.
 On the one hand, the spectrum is practically discrete,
 confirming the picture suggested in \cite{6}, where aperiodic
 conductance oscillations are due to the superposition of
 incommensurate harmonics. On the other hand, the spectrum as a
 whole resembles discrete white noise, which is close in
 properties to continuous white noise.

\begin{figure}[h]
\centerline{\includegraphics[width=2.7 in]{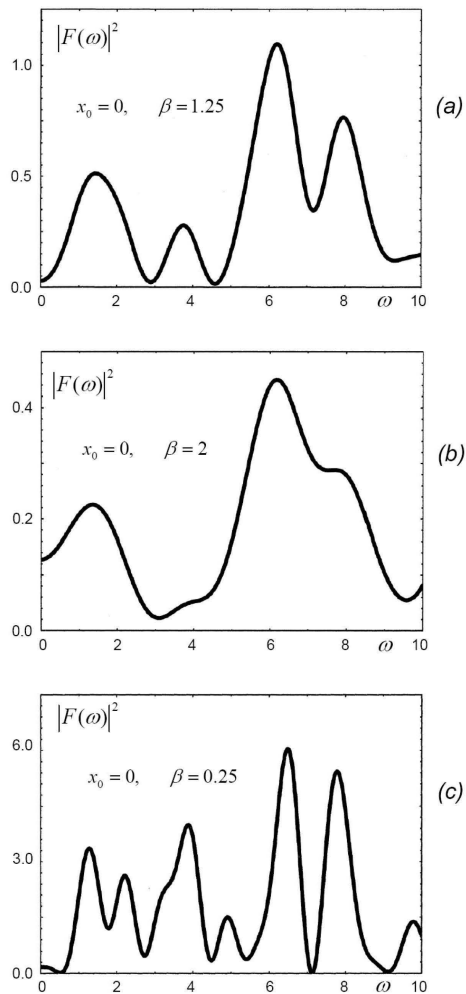}}
\caption{\small Dependence of the Fourier spectrum on the
choice of the smoothing function:  (a)  $\beta=1.25$,
(b) $\beta=2.0$, (c) $\beta=0.25$.
} \label{fig4}
\end{figure}

\begin{center}
{\bf 3. Dependence of results on the treatment
procedure}
\end{center}

Let consider
how results are influenced by the choice of
the smoothing function (9), which for $\mu=T\ln2$
leads to the line shape (11) and depends on one
parameter $\beta=1/T$. For the choice $\beta=1.25$ it
provides the small value $G(\mu_0)\approx 0.02$ at the
boundary of the working range and the spectrum of
$|F(\omega)|^2$ has a clearly
discrete character (Fig.\,4,a), which is
practically unchangeable in the interval $\beta=1.0\div 1.5$.
When $\beta$ is increased, the spectral lines are extended
in accordance with (11) and their partial confluence occurs
(Fig.\,4,b).
If $\beta$ is diminished, then the value  $G(\mu_0)$ ceased to
be small and the abrupt cutoff is restored, resulting in the
appearance of the parasite oscillations and additional maxima
of $|F(\omega)|^2$ (Fig.\,4,c). The role of the smoothing
procedure is exactly in removing such oscillations, which
have no relation to the true spectrum.

In spite of the evident problems arising due to deviation
from the optimal regime of treatment, the frequencies of
the discrete harmonics reveal the surprising stability
under the change of $\beta$ more than the order of
magnitude (Fig.\,5), and there are no doubts in there objective
origin. The small $\beta$ dependence is related with a change
of the shape of spectral lines and their mutual influence by each
other.  In fact, Fig.\,5 demonstrates that approximation of
independent harmonics is working sufficiently good.

\begin{figure}[h]
\centerline{\includegraphics[width=3.2 in]{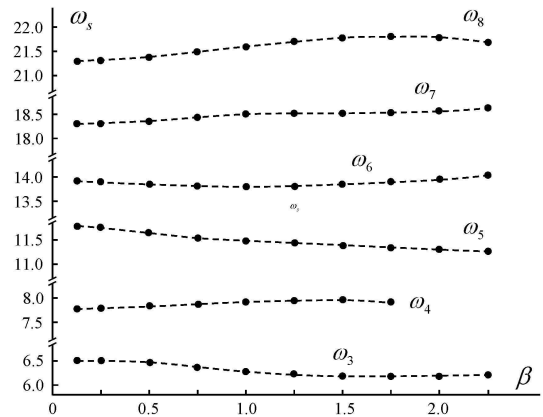}}
\caption{\small Frequencies of the most intensive
harmonics against the parameter $\beta=1/T$ determining
the form of the smoothing function.
} \label{fig5}
\end{figure}

If the experimental range of fields was smaller,
then the regime of the optimal line resolution (Fig.\,4,a)
might be absent, and Fig.\,4,b with confluent lines would be
changed immediately by Fig.\,4,c with parasite oscillations. One
can suggest, that at the present experimental conditions  the
line resolution is  also  incomplete, and they are subjected
to partial confluence.

\begin{center}
{\bf 4. Continuous component of the spectrum}
\end{center}

The Fourier spectrum in Fig.\,3 looks purely discrete, and
it is slightly unnatural.  The analogy with $1D$ systems
leads to conclusion that the distribution  $P(\rho)$
undergoes systematic  variations of deterministic nature
leading to
the $\rho$ oscillations in the specific sample.
However, analogous oscillations (of the random character)
should take place even for the stationary distribution
$P(\rho)$ due to its finite width. It would be more natural,
if Fig.\,3 contained the continuous component
and the discrete lines were observed
against its background. In fact, the continuous component exists
actually, and below we try to estimate it.

\begin{figure}[h]
\centerline{\includegraphics[width=3.2 in]{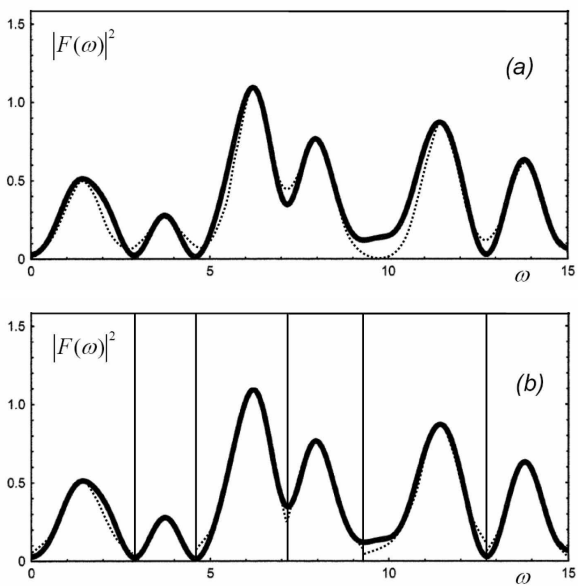}}
\caption{\small (a) Comparison of the experimental
Fourier spectrum (solid line) with relation (8) (dotted
line); frequencies $\omega_s$ and amplitudes $|A_s|$
were determined by the positions and heights of
maxima in Fig.\,3. (b) Fitting the form of separate
lines according to Eq.\,18.
}
\label{fig6}
\end{figure}

According to (8),
in the approximation of independent harmonics
the spectrum of $|F(\omega)|^2$  is represented as a
superposition of functions $g^2(\omega-\omega_s)$, whose
form is known beforehand. The test of relation (8) is
presented in Fig.\,6,a, where frequencies  $\omega_s$ and
amplitudes $|A_s|$ were determined by the positions and heights
of maxima in Fig.\,3; no fitting was made in respect of the
form of $g^2(\omega-\omega_s)$. Agreement looks satisfactory, but
not complete: the line widths can differ from the theoretical
predictions both in the greater or smaller side, while
the observed shape of lines is not always symmetric.
It is natural to relate these facts with
existence of the continuous component of the spectrum. If
this component is slowly variated, then counting $\omega$
from the line center, one can accept
$$
F(\omega)=A g(\omega)+B \,,
\eqno(14)
$$
where $B$ is constant on the scale of the line width.
Then
$$
|F(\omega)|^2=c_1 g^2(\omega)+c_2 g(\omega)+c_3 \,,
\eqno(15)
$$
where
$$
c_1=|A|^2\,,\quad c_2=2|A||B|\cos{\chi}\,,\quad
c_3=|B|^2
\eqno(16)
$$
and $\chi$ is determined by the difference of the
$A$ and $B$ phases. If $g(\omega)$ is normalized to unity at
$\omega=0$, then $g^2(\omega)$ corresponds to the more
narrow maximum than $g(\omega)$; hence, one has the broadening
of the line for $c_2>0$ and narrowing for $c_2<0$, while
the asymmetric form arises
if $B$ changes essentially near the maximum.
According to (15), $|F(\omega)|^2$ is determined by a
superposition of three basis functions $g^2(\omega)$, $g(\omega)$
and $1$, whose coefficients
may be determined by minimization of rms
deviation. This is a standard fitting procedure \cite{9},
which is linear and unique. However, practically it leads
to nonphysical results due to violation of the condition
$$
|c_2|\le 2 \sqrt{|c_1||c_3|} \,,
\eqno(17)
$$
following from (16). It looks that the optimal
fitting corresponds to the limiting values $\pm 1$ for
$\cos{\chi}$; in this case, parameters $A$ and
$B$ can be considered as real\,\footnote{\,It should
be stressed that reality of $A$ and $B$ is only
effective: in fact, their phases are correlated, being
equal to each other or differing by $\pi$. It looks that
the phases of discrete harmonics are adjusted to a
specific realization of the continuous component. The mechanism
of this phenomenon is not clear and should be considered as
an experimental fact.}, which leads to the equation
$$
|F(\omega)|^2=\left[ (F_0-B) g(\omega)+B\right]^2 \,,
\eqno(18)
$$
where $F_0=|F(0)|$. We have taken into account that the given
procedure is reasonable near the maximum of $|F(\omega)|^2$,
and its position is naturally kept fixed.
Fitting the form of each line according to Eq.\,18, we come to
Fig.\,6,b:  agreement is practically ideal for the most of lines,
while its absence for some of them is probably related with
existence of secondary harmonics which were hidden
against the background of main lines.

When the coefficients $c_1$, $c_2$, $c_3$ are
determined, one can exclude contributions proportional to
$g^2(\omega)$ and $g(\omega)$ from $|F(\omega)|^2$.
Producing this procedure for all of lines, one  comes
to the "residual" spectrum presented in Fig.\,7. Abrupt maxima in
this spectrum are probably related with the secondary
discrete harmonics, while the rest part is naturally
ascribed to the continuous spectrum; its density corresponds
to  $10-15\%$ of the main lines intensity, and the origin of
its smallness is discussed in Sec.\,5. The continuous
component approximately corresponds to the white noise spectrum
for $\omega\alt 2\pi/\xi_B$, but the suggested slowness of
its variations is not confirmed;
hence the obtained result
should be considered only as a rough estimate.\,\footnote{\,The
treatment procedure is not unique
for asymmetric lines, since one can fit their right or left
part. This ambiguity was used in order to avoid the appearance
of nonphysical negative values for $|F(\omega)|^2_{res}$. }

\begin{figure}
\centerline{\includegraphics[width=3.0 in]{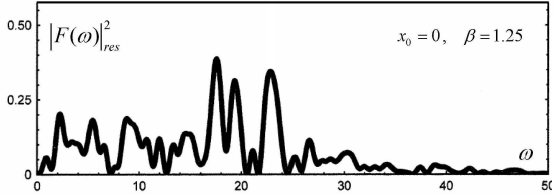}}
\caption{\small The residual spectrum obtained by
exclusion of contributions of the main discrete frequences.
The abrupt maxima are probably related with the secondary
harmonics, which were hidden
against the background of the main
spectral lines. The rest is naturally ascribed to the
continuous component.
} \label{fig7}
\end{figure}

Strictly speaking, the quantity $B$ in Eq.\,15 represents not only
the continuous component, but also the contributions from the
neighboring lines. It is not essential for correct estimation
of the amplitude $A$, which is confirmed by  analogous
calculation for the smoothing function with $\mu=2$, $T=0.5$:
the  tails  of latter are essentially different from
those of  function (11), while the result is not very different
from Fig.\,7.

It should be noted, that the spectrum $|F(\omega)|^2$ in Fig.\,3
looks \mbox{"more discrete"}, than it is such in reality:
it is related with the fact
that the quantity $B$, being approximately constant
in the vicinity of each maximum, changes its sign in the gaps
between certain lines. The residual spectrum in Fig.\,7 does not
coincide with the quantity $B$, but contains the errors of
fitting and effects of interference between neighboring lines.

\begin{figure}[h]
\centerline{\includegraphics[width=2.8 in]{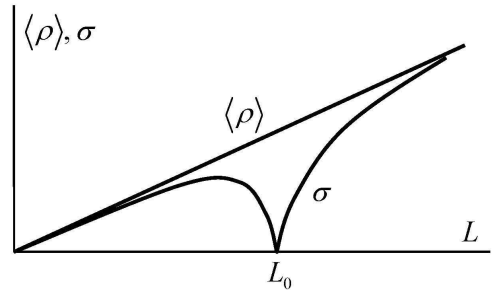}}
\caption{\small Behavior of $\langle\rho\rangle$ and
$\sigma$ in the metallic regime according to Eq.\,19 for the
initial condition (22). For the choice $\rho_0=\langle\rho\rangle$
accepted in the figure, dependence of $\langle\rho\rangle$
against $L$ remains the same as for the initial condition
(20). The picture does not change qualitatively, if
a typical value from distribution (21) is chosen for $\rho_0$.
} \label{fig8}
\end{figure}

\begin{center}
{\bf 5. Why the spectrum is practically discrete?}
\end{center}

Let us discuss the reasons for smallness of the continuous
component of the spectrum. According to \cite{32}-\cite{37},
the evolution of distribution $P(\rho)$ in 1D systems is
described by the diffusion type equation
$$
\frac{\partial P(\rho)}{\partial L} =
\alpha\,\frac{\partial}{\partial \rho}
\left[\,\rho(1\!+\!\rho)\,\frac{\partial P(\rho)}{\partial \rho}
\,\right] \,, \eqno(19)
$$
where $\alpha L$ plays a role of time. This equation is obtained
in the random phase approximation, which is adequate in the
quasi-metallic regime, i.e. in the deep of the allowed band for
weak disorder \cite{6}.  The natural initial condition for
(19) has a form
$$
P(\rho)=\delta(\rho)\,\quad \makebox[1.5cm]{\rm for}  L=0\,,
\eqno(20)
$$
since for the zero length of the system its resistance is zero
independently of the random potential realization. Such
initial condition leads to the distribution
$$
P\left(\rho\right) = (\alpha L)^{-1}
\exp\left\{-\rho/\alpha L\right\}
\eqno(21)
$$
for small $L$ (when typical values of $\rho$ are small) and to
the log-normal distribution at large $L$ (when typical values of
$\rho$ are large). The mean value $\langle\rho\rangle$ for
distribution (21) coincide with rms deviation $\sigma$, while
$\sigma$ grows faster than $\langle\rho\rangle$ in the log-normal
regime. Let consider the more general initial condition
$$
P(\rho)=\delta(\rho-\rho_0)\,\quad \makebox[1.5cm]{\rm for}
L=L_0\,,
\eqno(22)
$$
whose meaning is discussed below.  Solution of Eq.\,19 with the
initial condition (22) is close to the Gaussian
distribution for $L$ close to $L_0$ (see Appendox 1), while
distribution (21) or the log-normal distribution is restored
for large $L$ where the finiteness of $\rho_0$ and $L_0$
becomes inessential. When distribution $P(\rho)$ is close to the
Gaussian one, it is reasonably described by two first moments,
whose evolution is easily obtained (see Appendix 1): the typical
situation is presented in Fig.\,8. It is easy to see that
$\langle\rho\rangle$ is essentially greater than $\sigma$
in the interval around $L_0$, whose width is of order $L_0$.

Let discuss the meaning of the initial condition (22).
Suppose we are measuring the resistance $\rho$ of the system
on the length $L_0$, creating different impurity
configurations\,\footnote{\,It can be made practically by
heating the sample till sufficiently high temperature.};
with sufficiently large number of configurations we shall
reproduce distribution (21). Let now change the procedure
and  select only configurations, whose resistance
$\rho$ falls in the small interval around $\rho_0$:
it artificially creates the ensemble with the narrow
distribution of type (22), whose evolution leads to the
picture presented in Fig.\,8. Let now take a single sample
with resistance $\rho_0$ at the length $L_0$. Dependence
$\rho(L)$ for this sample can be described theoretically,
if all details of the impurity configuration are known.
Usually such information is absent, and only general statistical
properties of the random potential are available.
In this case one can establish only an approximate corridor
for possible dependencies $\rho(L)$: this corridor is illustrated
by Fig.\,8.

Equation (19) is obtained in the random phase
approximation, which eliminates all oscillation effects.
Fortunately, evolution of $\langle\rho\rangle$ may be
investigated exactly without any assumptions (see Appendix 2).
In the quasi-metallic regime and for the "natural" ideal leads
\cite{6} the following result is valid
$$
\langle\rho\rangle=\rho_0+ \frac{1\!+\!2\rho_0}{2}
\left( {\rm e}^{2\epsilon^2 l} -1\right) +
\eqno(23)
$$
$$ +
\frac{\epsilon^2}{\delta} \sqrt{\rho_0(1\!+\!\rho_0)}
\left[\,  {\rm e}^{2\epsilon^2 l} \sin{\psi} -
{\rm e}^{-\epsilon^2 l} \sin{(2\delta l+\psi)} \right]\,,
$$
which was obtained for the discrete Anderson model;
here $\delta=k_F a_0$, $\epsilon^2=W^2/4\delta^2$,
$l=(L-L_0)/a_0$, $k_F$ is the Fermi momentum, $a_0$ is the
lattice constant, $W$ is the amplitude of the random
potential, and $\psi$ is determined by the difference of phases
entering the transfer matrix
specified at the scale $L_0$. One can see the existence of
oscillations, whose period is determined by the de Brougli
wavelegth ($2\delta l = 2 k_F L$  for $L_0=0$); their
amplitude can be comparable with
$\rho_0\ll 1$ in spite of the small parameter $\epsilon^2/\delta$.
The quantity $\psi$ is completely stochastizated in the random
phase approximation, and averaging over it eliminates
oscillations and restores the result following from Eq.\,19
(see $(A.6)$ in Appendix 1). Beyond the metallic regime
($\epsilon^2\agt \delta$) the amplitude of oscillations
is certainly exceeding $\rho_0$,
$$
\langle\rho\rangle=\rho_0+
\frac{1}{3}\left( \frac{\epsilon^2}{\delta} \right)^{2/3}
\left( \frac{1\!+\!2\rho_0}{2} -\cos{\psi}
\sqrt{\rho_0(1\!+\!\rho_0)}\right)\cdot
$$
$$
\qquad       \cdot
\left[\,  {\rm e}^{x_1 l}  -
2\,{\rm e}^{-x_1 \!l/2} \cos{ \left(  \frac{\sqrt{3} x_1
l}{2}+\frac{\pi}{3} \right) } \right]\,, \
\eqno(24)
$$
and they do not disappear after averaging over $\psi$
due to fundamental inapplicability of the random phase
approximation \cite{6}
(here  $x_1=\left( 8\epsilon^2 \delta^2\right)^{1/3}$).
The amplitude of oscillations in the metallic regime increases
essentially, when the foreign ideal leads are used,
$$
\langle\rho\rangle=\rho_0+ \frac{1\!+\!2\rho_0}{2}
\left[-1+\Delta_2^2\, {\rm e}^{2\epsilon^2 l} -\Delta_1^2 \,
{\rm e}^{-\epsilon^2 l} \cos{2\delta l}
\right] +
$$
$$ +
\Delta_1 \sqrt{\rho_0(1\!+\!\rho_0)}\cdot
\left\{ \, {\rm e}^{2\epsilon^2 l}\,\Delta_2 \cos{\psi}-
\right. \qquad
\eqno(25)
$$
$$
\qquad    -\left.
{\rm e}^{-\epsilon^2 l} \left[ (\Delta_2\!-\!1)
\cos{\psi}\,\cos{2\delta l} + \cos{(2\delta l+\psi)}
\vphantom{\Delta_2^2}\right]
\right\}\,,
$$
where $\Delta_1$ and $\Delta_2$ are defined in Appendix 2 and can
be large. Thereby, under rather general conditions the amplitude
of the $\langle\rho\rangle$ oscillations is comparable
with $\rho_0$ and in the vicinity of $L_0$ is certainly greater
than $\sigma$. Unfortunately, results of type (23--25)
for higher moments are practically inadmissible due to
tremendous calculations, and only general arguments can be
given. It is natural to expect, that the second moment
$\langle\rho^2\rangle$ is oscillating with the amplitude of
the order of  $\rho^2_0$, which exceeds essentially
the quantity $\sigma^2$.
Hence,  oscillations of  the distribution width
due to systematic variations exceed essentially the width of the
distribution in the absence of oscillations. As a result,
oscillations of higher moments are also essential.

Applicability of the above analysis to the situation under
consideration is determined by the fact that the range of
fields $B=2\div 6 T$ was effectively treated
for the smoothing function with $\mu=2$, $\beta=2$ used in
\cite{11a}, which corresponds to variation of field
by a factor of three. Since  $L\propto B^{-1/2}$,
the corresponding variations of $L$ occur\,\footnote{\,The
length $L$ is estimated as a value of the
$x$ coordinate, for which the quadratic potential
$m \omega_B^2 x^2 $ \cite{7} becomes of the order of the Fermi
energy $\epsilon_F$, so $L\propto B^{-1}$. However, if
$\epsilon_F$ is comparable with the first Landau level,
then it should be replaced by $\hbar \omega_B =\hbar e B/mc$,
which gives
$L\propto B^{-1/2}$.  As clear from Sec.10,  the latter
estimate is more adequate near the middle of the experimental
range of magnetic fields.}
 within a factor 1.7.
If $L_0$
is chosen in the middle of the interval, then deviations from
$L_0$ are on the level of 30\%. These deviations are even smaller
for the smoothing function with $x_0=0$, corresponding to
(11).  The situation does not change, if the experimental range
of fields is increasing, because the choice of the smoothing
function is determined by the same considerations:  only $\mu_0$
value becomes different, while all proportions in Fig.\,2 remain
unchanged.

\begin{center}
{\bf 6. Shift oscillations}
\end{center}

Returning to formulas of Sec.\,2, one can see that
Eq.\,7 is more general that Eq.\,8: the former is an exact consequence
of Eq.\,6, while the latter suggests weak intersection of
spectral lines. One can hope that a treatment on the base of Eq.\,7
with the use of representation (14) and a separate fitting of
the real and imaginary part of $F(\omega)$ provides
the more smooth form for $|F(\omega)|^2_{res}$. Unfortunately,
these hopes are not confirmed: the obtained picture is not very
different from Fig.\,7 and does not prove
the more complicated
treatment. However, study of the real and imaginary parts of
$F(\omega)$ reveals a lot of interesting aspects, which are
discussed in the present and subsequent sections.

\begin{figure}[h]
\centerline{\includegraphics[width=2.7 in]{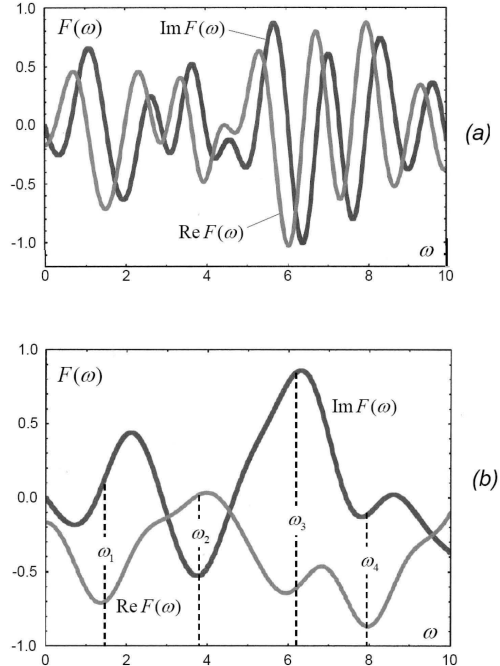}}
\caption{\small (a) Behavior of the real and
imaginary parts of $F(\omega)$ indicates the existence of
the shift oscillations ${\rm e}^{i\omega a}$ with $a=4.3$;
 (b) Behavior of ${\rm Re}F(\omega)$ and ${\rm Im}F(\omega)$
after exclusion of the shift oscillations. Dashed lines indicate
the positions of the $|F(\omega)|^2$ maxima.
} \label{fig9} \end{figure}

Functions ${\rm Re}F(\omega)$ and ${\rm Im}F(\omega)$
appear to be quickly oscillating  (see Fig.\,9,a). The reason
of the oscillations is easy to understand: if $F(\omega)$  is
the Fourier transform of $f(x)$, then a shift of the $x$
origin leads to the correspondence
$$
f(x-a) \,\,\Longleftrightarrow \,\, {\rm e}^{i\omega a} F(\omega)
\eqno(26)
$$
and the quick oscillations arise for large $a$. One can puzzle
himself by a question on the "natural" choice of the $x$
origin, which corresponds to slow variations of the Fourier
transform. Estimating the average period of oscillations
in Fig.\,9,a, one obtains
$a=4.3$ and
elimination of the factor $\exp(i\omega a)$ leads to
Fig.\,9,b, where ${\rm Re}F(\omega)$ and  ${\rm Im}F(\omega)$
change on the same scale as $|F(\omega)|$. The obtained value
of $a$ signifies that the conductance dependence against
the magnetic field has a natural origin $B_0=8.3 T$, which
lies beyond the upper boundary of the experimental range. Since
large fields correspond to small system lengths, one can suggest
the following interpretation.

When we study the evolution of the distribution $P(\rho)$ against
the length $L$ of a 1D system, the natural origin is evidently
$L=0$.  However, if evolution begins with a finite scale $L_0$,
then a form of the distribution at large $L$ appears to be the
same as in the case $L_0=0$.  If we consider a situation at large
$L$ and try to extrapolate to the initial stages of evolution,
then we cannot establish from which scale it is
initiated\,\footnote{\,Such extrapolation is complicated by the
factors, shifting the $L$ origin (see Sec.\,5 in  \cite{6}).}: we
can only claim that this scale is small in comparison with those
under consideration. In terms of the magnetic field it means
that evolution begins with a certain large value $B_0$. However,
due to nonlinearity of the relation between $L$ and $B$  the
extrapolated $B_0$ value appears to be not very
large\,\footnote{\,As clear from Sec.5, the main contribution to
$F(\omega)$ occurs from the middle of the experimental
range of fields, where the relation between $L$ and $B$
is practically linear. Correspondingly, extrapolation to the
origin is also effectively linear.}.


Suggested interpretation looks rather logical and provides
indirect confirmation of the analogy with 1D systems.
The formal arguments on the choice of the natural origin are
presented in Appendix 3.

\begin{center}
{\bf 7. The phase distribution}
\end{center}

When the natural origin of the argument of $f(x)$ is established,
the information on the real and imaginary parts of $F(\omega)$ may
be used for a subsequent analysis. Estimating values of
${\rm Re}F(\omega)$ and ${\rm Im}F(\omega)$ at the points of
maximum for $|F(\omega)|^2$, one can determine the
complex phases of the coefficients $A_s$ entering Eqs.\,6--8.
Their distribution is illustrated in Fig.\,10.

\begin{figure}[h]
\centerline{\includegraphics[width=2.8 in]{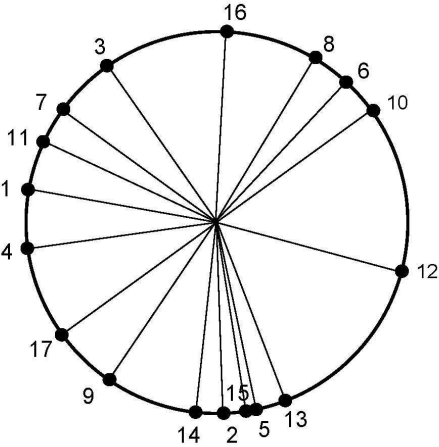}}
\caption{\small Distribution of the phase factors
 ${\rm e}^{i\varphi_s}$ on the unit circle; numbers near
 points indicate the corresponding $s$ value.
} \label{fig10}
\end{figure}

As was discussed in Sec.\,2, the discrete white noise is
analogous to the continuous one, if phases of $A_s$
are completely random. According to Fig.\,10, their distribution
is sufficiently uniform and does not contradict to their
expected
randomness.
It would be interesting to verify, how the
phase distribution changes
when the new impurity configurations are created.

\begin{center}
{\bf 8. Evidence of the exponential growth} \end{center}

According to Eq.\,7 contributions of the discrete harmonics to
$F(\omega)$ are proportional to $g(\omega-\omega_s)$, and
should lead to extrema
of the real and imaginary parts of $F(\omega)$
at the points $\omega_s$.  However, Fig.\,9,b demonstrates that
extrema of ${\rm Re}F(\omega)$ and ${\rm Im}F(\omega)$ are
realized at different points, not coinciding
with maxima of $|F(\omega)|^2$. It indicates invalidity of Eq.\,7
and casts doubt on the initial expression (6), from which it was
derived.

Let recall that according to Eq.\,2 amplitudes of
oscillations are not fixed,
but subjected to the exponential growth. If the
latter is taken into account, then Eq.\,6 is modified as
$$
f(x)=\frac{1}{2}\sum_s
\left[ A_s {\rm e}^{i\omega_s x+\alpha_s x}
+A_s^* {\rm e}^{-i\omega_s x+\alpha_s x} \right] G(x)\,,
\eqno(27)
$$
and instead of (7) one comes to the result
$$
F(\omega)=\frac{1}{2} \sum_s
\left[ A_s g(\omega+\omega_s-i\alpha_s)
+A_s^* g(\omega-\omega_s-i\alpha_s) \right]   \,.
\eqno(28)
$$
Let concentrate on the contribution of the single harmonics
$\omega_s$ and accept, shifting the $\omega$ origin
$$
 A=A'+i A''=|A|{\rm e}^{i\varphi}\,,\qquad
 g(\omega-i\alpha)=g_1(\omega)+i g_2(\omega)
 \,. \eqno(29)
$$
If $\omega_1$ and $\omega_2$ are positions of extrema for
${\rm Re}F(\omega)$ and ${\rm Im}F(\omega)$, then the
following relations are valid
$$
A' g'_1(\omega_1)-A'' g'_2(\omega_1)=0\,,
$$
$$
A' g'_2(\omega_2)+A'' g'_1(\omega_2)=0\,,
\eqno(30)
$$
and the phase $\varphi$ of the coefficient $A$ is determined
by the condition
$$
\tan\varphi= \frac{g'_1(\omega_1)}{g'_2(\omega_1)}=
-\frac{g'_2(\omega_2)}{g'_1(\omega_2)}\,.
\eqno(31)
$$
If the experimental values of $\omega_1$ and  $\omega_2$ are
known, then $\varphi$ and $\alpha$ are determined by the
following algorithm.  According to Eq.\,29,  $g_1(\omega)$ and
$g_2(\omega)$ depends on $\alpha$, and  $g_2(\omega)\to 0$ for
$\alpha\to 0$; hence for small $\alpha$ the former fraction in
(31) is large, while the latter is small in modulus. Increasing
$\alpha$, one can reach equality in Eq.\,31, which determines
$\tan\varphi$ and $\alpha$. The known value of tangent
specifies $\varphi$ to additive contributions multiple of $\pi$,
and to establish the correct quadrant for $\varphi$ one can use
the relations
$$
A' g_1(\omega_1)-A'' g_2(\omega_1)=F_1\,,
$$
$$
A' g_2(\omega_2)+A'' g_1(\omega_2)=F_2\,,
\eqno(32)
$$
where $F_1$ an $F_2$ are values of ${\rm Re}F(\omega)$
and ${\rm Im}F(\omega)$ in the corresponding extrema.

If the function $g(\omega)$  in Eq.\,11 is normalized to unity
at $\omega=0$, then setting
$$
z=\pi T\omega\,, \qquad \gamma=\pi T \alpha\,,
\eqno(33)
$$
one has
$$
g_1(\omega)=\frac{\cos{\!\gamma}\, z \sinh z
                +\gamma\sin{\!\gamma} \cosh z}
{ \sinh^2 z +\sin^2{\!\gamma}} \,,
$$
$$
g_2(\omega)=\frac{\sin{\!\gamma} \, z \cosh z
-\gamma\cos{\!\gamma}\, \sinh z }
{ \sinh^2 z +\sin^2{\gamma}} \,,
\eqno(34)
$$
$$
|g(\omega+i\alpha)|^2=\frac{z^2+\gamma^2}
{ \sinh^2 z +\sin^2{\!\gamma}} \,.
$$
It is easy to see that extrema of $g_1(\omega)$ and
$|g(\omega+i\alpha)|$ are realized at $\omega=0$, i.e.
they are  not affected by finiteness of $\alpha$, so
it is natural to count $\omega_1$ and $\omega_2$ from
zero. The function $g_2(\omega)$ is odd and its small
addition to $g_1(\omega)$ shifts the extremum to right
or left, in dependence on the sign of such addition.
In the case of functions ${\rm Re}F(\omega)$ and ${\rm
Im}F(\omega)$, the additions to $g_1(\omega)$,
proportional to $g_2(\omega)$, have the opposite signs and
provide different signs for $\omega_1$ and $\omega_2$.
For small $\alpha$ one can simplify Eq.\,34, retaining
the first order in $\gamma$,  and use the smallness of
$\omega_1$ and $\omega_2$:
%
%
$$
\tan\varphi= - \frac{\omega_1}{\alpha}=
\frac{\alpha}{\omega_2}\,,\qquad \alpha^2=-\omega_1\omega_2
\,.
\eqno(35)
$$
In this approximation Eq.\,32 gives  $A'\approx F_1$,
$A''\approx F_2$, which allows to establish the sign of $\alpha$
and choose the correct quadrant for $\varphi$, accepting that
the latter belongs to the interval $(-\pi, \pi)$:
$$
\alpha=\sqrt{-\omega_1\omega_2}\, {\rm sign}(\omega_2 F_1 F_2)\,,
$$
$$
\varphi= -\arctan(\omega_1/\alpha)+
\pi\, {\rm sign}F_2 \,(1-{\rm sign}F_1)/2\,.
\eqno(36)
$$
It should be stressed that nontrivial information
contained in $|\tan\varphi|$ is determined by
$\omega_1$ and $\omega_2$, while $F_1$ and $F_2$ are
used only for the choice of the correct quadrant.
Estimation (36) for $\varphi$ differs radically from the
estimate given in Sec.\,7 and coincides with it only in the
case, if the shifts of extrema of ${\rm Re}F(\omega)$
and ${\rm Im}F(\omega)$ are actually related with the
exponential growth.

Comparison of two estimations of $\varphi$  (see Table)
demonstrates their approximate agreement for all
harmonics and confirms the suggested
mechanism for shifts of extrema. Small differences
between two estimations may be related with the
approximate character of formula (36), and with
other factors, such as
mutual influence of harmonics, existence of the
continuous spectrum, experimental inaccuracies, etc.

  \begin{center}
{\it Table.} \quad  Estimations of phases\\
obtained in Sec.\,7 ($\varphi_s$) \\
and according formula (36)   ($\tilde\varphi_s$).
  \vspace{4mm}

\begin{tabular}{||c|c|c||c|c|c||}
\hline
$s$      &  $\varphi_s$ & $\tilde\varphi_s$ &
  $s$      &  $\varphi_s$ & $\tilde\varphi_s$ \\
  \hline
    &  & &    &  &  \\
1  & $ 170^\circ$    & $ 161^\circ  $ &
  10 & $  37^\circ $   & $ 37^\circ  $  \\
2  & $ -88^\circ $   & $ -84^\circ  $ &
  11 & $  155^\circ $  & $ 145^\circ  $ \\
3  & $ 125^\circ $   & $ 120^\circ  $ &
  12 & $ -15^\circ $   & $ -20^\circ  $ \\
4  & $ -172^\circ $  & $ -162^\circ $ &
  13 & $  111^\circ $  & $ 121^\circ  $ \\
5  & $ -78^\circ $   & $ -67^\circ  $ &
  14 & $  -96^\circ $  & $ -106^\circ $ \\
6  & $ 47^\circ $    & $ 40^\circ   $ &
15 & $  -81^\circ $  & $ -72^\circ  $ \\
7  & $ 144^\circ $   & $ 153^\circ  $ &
16 & $  87^\circ $   & $ 84^\circ   $ \\
8  & $ 59^\circ $    & $ 49^\circ   $ &
17 & $ -144^\circ $  & $ -140^\circ $ \\
9  & $ -124^\circ $  & $ -153^\circ $ &   &  & \\
    &  & &    &  &  \\
 \hline
 \end{tabular}
\end{center}
  \vspace{6mm}

\begin{center}
{\bf 9. Distribution of the growth exponents and the
frequency differences }
\end{center}

Let give a summary of theoretical results for
evolution of the $\rho$ moments. Accepting that $\langle \rho^n
\rangle$ behaves with $L$ as $\exp{xL}$, we can
obtain the algebraic equation of $(2n\!+\!1)$th order for the
grow exponent $x$. For $n=1$ and $n=2$ such equations can be
given explicitly \cite{6}
$$
x\left(x^2 +4{\cal E}\right)= 2W^2  \,,
$$
$$
x\left(x^2+{4\cal E} \right)\left(x^2 +16{\cal E}\right)=
42W^2 x^2+96 W^2 {\cal E}  \,,
\eqno(37)
$$
where $\cal E$ is the energy counted from the lower
edge of the initial band, and $W$ is the amplitude of the
random potential.

The structure of equations for arbitrary $n$th moments can be
 established using argumentation presented in Section 4 in
 \cite{6}. In the deep of   the allowed and forbidden bands, only
 diagonal elements can be retained in matrices (43) and (47)  in
 \cite{6} and their analogs  for higher moments. As a result,
 we arrive at the equation
$$
\prod_{k=0}^{2n}  \left[x-2(n\!-\!k)\delta-B_n^k
\epsilon^2\right] = O(\epsilon^4 \delta^{2n-1})\,,
\eqno(38)
$$
where $\epsilon^2=W^2/4{\cal E}$, $\delta^2=-{\cal E}$,
$B_n^k=n(2n\!-\!1)+3k(k\!-\!2n)$.
A similar equation near the band edge
$$
x^{2n+1}=\sum\limits_{k=0}^{k_{max}} C_k W^{2k} x^{2n+1-3k}\,,
\quad k_{max}=\left[\frac{2n+1}{3} \right]
\eqno(39)
$$
follows from observation that all terms of the equation
have the same order of magnitude at $x\sim
\delta\sim\epsilon^2$ and only combinations
$\delta^{2n}\epsilon^{2m}$ with $n\ge m$ are allowed, among
which only $\delta^{2n}\epsilon^{2n}\sim W^{2n}$ remain finite
at $\delta\to 0$.  Nontrivial roots of Eq.\,39 are of the order of
$W^{2/3}$ and leads to the incommensurate oscillations in Eq.\,2.

\begin{figure}
\centerline{\includegraphics[width=2.5 in]{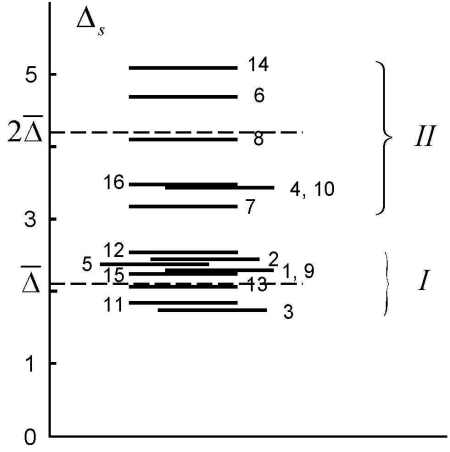}}
\caption{\small Distribution of differences
$\Delta_s=\omega_{s+1}-\omega_s$ between
the neighboring frequencies; the $s$ value is indicated
near the corresponding lines. Differences $\Delta_s$
break up into two groups $I$ and $II$, localized near
$\bar\Delta$ and $2\bar\Delta$.
}
\label{fig11}
\end{figure}

In the deep of the allowed band parameters $\delta$ and
$\epsilon$ are complex, and one should make a replacement
$\delta\to i\delta$,  $\epsilon\to -i\epsilon$ in order
to reduce them to the real form.  Then Eq.\,38 gives the
complete set of exponents for the extremely metallic regime
($\epsilon^2\ll \delta$)
$$
x^k_n=2i(n\!-\!k)\delta-B_n^k
\epsilon^2 + O(\epsilon^4/\delta)\,,\quad k=0,1,\ldots,2n\,.
\eqno(40)
$$
One can see that all frequencies of oscillations in Eq.\,2
are integer multiple of the quantity $\bar\Delta=2\delta$,
i.e. their incommensurability disappears. In the ideal case
all differences of the neighboring frequencies
$\Delta_s=\omega_{s+1}-\omega_s$ should be equal to
$\bar\Delta$. Practically
certain harmonics are not manifested due
to their weak intensity, so differences $\Delta_s$ are
\mbox{"quantized",} and may be equal  $\bar\Delta$,
$2\bar\Delta$, $3\bar\Delta$, etc.
\begin{figure}[h]
\centerline{\includegraphics[width=1.7 in]{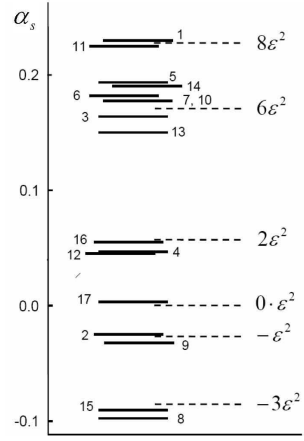}}
\caption{\small Distribution of the growth
exponents $\alpha_s$ and their comparison with succession
(41) for the proper choice of $\epsilon^2$.  The
$s$ value is indicated near the corresponding  lines.
}
\label{fig12}
\end{figure}

The distribution of differences $\Delta_s$ for 17 harmonics,
evident from Fig.\,3, is represented in Fig.\,11. They
break up into two groups $I$ and $II$, localized near
$\bar\Delta$ and $2\bar\Delta$, where $\bar\Delta$ is chosen from
the best agreement. Absence of exact quantization should not
cause any trouble, since
it refers only to the extremal
metallic regime. In actuality the metallic regime is not extremal
and the corrections indicated in Eq.\,40 become visible.

The growth exponents $\alpha_s$ are determined by the
real part of Eq,40 and are integer multiple of the
quantity $\epsilon^2$. Going over all possible values of
$n$ and $k$, one obtains the infinite succession of
exponents,
which looks as follows near the origin:
$$
\ldots,
\,\,-3\epsilon^2, \,\, -\epsilon^2, \,\,0, \,\,2\epsilon^2,
\,\,3\epsilon^2, \,\,6\epsilon^2, \,\,8\epsilon^2, \,\,\ldots
\eqno(41)
$$
The exponents obtained by the treatment of the
experimental data according to Eq.\,36 are represented
in Fig.\,12 with the opposite sign\,\footnote{\,Exponents
$\alpha_s$ change their sign in the course of transition from
the magnetic field $B$ to the effective system length $L$ (the
left and right direction trade their places). When passing
from conductance to resistance, there is no change of
the $\alpha_s$ signs, since the small fluctuations of two
quantities are proportional to each other.}
and reproduce this succession rather well for the
proper choice of $\epsilon^2$. The only exclusion is
absence of the term $3\epsilon^2$, which is probably
related with low intensity of the corresponding spectral lines.
Manifestations of only exponents close to zero  is naturally
explained by the fact that harmonics with large (in modulus)
exponents are localized near the ends of the experimental range
of magnetic fields and are invisible in its middle.

It should be noted that the maximal exponent for the
moment $\langle \rho^n \rangle$ in the forbidden band is realized
for $k=2n$,
$$
x^{max}_n= 2n\delta+n(2n-1)\epsilon^2\,,
\eqno(42)
$$
and in the allowed band for  $k=n$,
$$
x^{max}_n= n(n+1)\epsilon^2\,.
\eqno(43)
$$
These results are in agreement with the functional form for
the log-normal regime
$$
x^{max}_n= an +b n^2/2
\eqno(44)
$$
with parameters $a$ and $b$, obtained in \cite{6}:
$a=2\delta-\epsilon^2$, $b=4\epsilon^2$  in the forbidden band,
and $a=\epsilon^2$, $b=2\epsilon^2$ in the allowed band.
\vspace{6mm}

\begin{center}
{\bf 10. Microscopical picture}
\end{center}

It is easy to see that quantization for the growth
exponents $\alpha_s$ (Fig.12) is more exact than for
the frequency differences $\Delta_s$ (Fig.11). This fact
allows a simple explanation.

\begin{figure}
\centerline{\includegraphics[width=3.0 in]{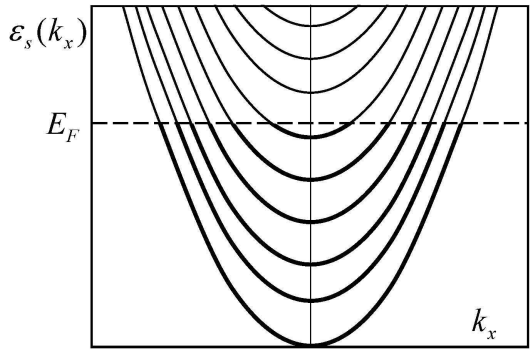}}
\caption{\small The transverse movement in a thin wire
is quantized, so $N_0$ discrete levels arise, which becomes 1D
subbands, if the longitudinal movement is taken into account. The
main contribution to the conductance oscillations is given by the
upper filled subband.
} \label{fig13}
\end{figure}

A thin wire is a quasi-1D system, in which
the transverse motion is quantized, so  $N_0$ discrete levels
$\epsilon_s^0$ arise. If the longitudional movement along the
$x$ axis is taken into account, the levels transform to the 1D
subbands with the spectra

$$
\epsilon_s(k_x)=\epsilon_s^0 + k_x^2/2m\,,
\eqno(45)
$$
whose states are filled below the Fermi level $E_F$
(Fig.13). The magnetic field $B$ affects most strongly
the upper filled subband with the minimal Fermi energy
$\epsilon_F$, restricting the movement in it by the
length $L$ determined by the condition
$$
m\omega_B^2 L^2 \sim \epsilon_F \sim E_0/N_0\,,
\eqno(46)
$$
so
$$
L \sim a \frac{B_0}{B \sqrt{N_0}} \,,
\eqno(47)
$$
where $E_0=\hbar^2/ma^2$ and $B_0=\phi_0/a^2$ are
the atomic units
of the energy and the magnetic field, while $\phi_0=\pi
\hbar c/e$ is the flux quantum.
The system conductance is determined by the sum of the subband
conductances, whose oscillations are exponentially decreasing
for large $L$ (see below). The main contribution to oscillations
is given by the upper filled subband where the length $L$ is
minimal, while the neighbouring subbands also have a certain
influence. This influence is rather essential
for the frequency differences $\Delta_s$ and broadens their
distribution (Fig.11), since $\omega_s$  are
determined by the Fermi momentum  $k_F$, whose value is different
for different subbands. As for the growth exponents $\alpha_s$
(Fig.12), a situation is strikingly different for them. The Fermi
momenta are small for upper subbands, and the approximation of
slow particles is applicable for the scattering on
impurities\,\footnote{\,One should take into account a
difference of the actual system from the Anderson model
considered in \cite{6}. In the Anderson model, the metallic
regime corresponds to a large concentration of weak impurities,
which can be treated by perturbation theory. In the actual
system, the weak disorder is created by a small concentration of
strong impuriries, for which the slow particles approximation
can be used. This difference does not affect the
results for the exponents $\alpha_s$, since on the scale of a
wavelength the configuration of the random potential can be
changed in wide limits without influence on the large scale
properties of wave functions.}, so the scattering amplitude is
independent of a momentum. The growth exponents are directly
related with the scattering amplitude and do not depend on the
Fermi momenta of 1D subbands. As a result, the neighbouring
subbands do not affect the exact quantization,
so Fig.12 corresponds to a strictly 1D
system. Deviations from exact quantization are related only with
nonextremality of the metallic regime and experimental
inaccuraces.

The minimal discrete harmonics in Fig.3 approximately
corresponds to 4 oscilations for a change of the magnetic field
in the interval $1\div 10 T$. Its frequency is determined by the
de Brougli wavelength $\lambda\sim a\sqrt{N_0}$ in the upper
subband, while the number of oscillations in the interval
of fields from $B_{min}$ till  $10 B_{min}$ is given by the
estimate
$$
N_{osc} \sim B_0/N_0 B_{min} \,,
\eqno(48)
$$
following from (47), and setting $B_0\sim 10^4 T$, $B_{min}=
1 T$,  $N_{osc}= 4$, one obtains
$$
N_0 \sim 2.5 \cdot 10^3 \,.
\eqno(49)
$$
It is half of the number of atoms in the cross section of a
wire with 25 nm in diameter \cite{5}, and corresponds to a
half-filled 3D band. In this case $\lambda\sim 50 a$, and the
actual length range
$$
L = 20\div 200 a \,
\eqno(50)
$$
is within the wire lenfth 310 nm \cite{5}.

Above we accept the condition $\hbar \omega_B \ll \epsilon_F$,
which is in fact violated in the middle of the
experimental range of magnetic fields. For large fields
another estimate $L\sim a (B_0/B)^{1/2}$ is more adequate,
which follows from Eq.46 after replacement $\epsilon_F$
by $\hbar \omega_B$. As a result, the lower bound in Eq.50
shifts from $20 a$ to $30 a$, which is not essential for
the given estimates.

As clear from Eqs.23,25, the expression for
$\langle \rho\rangle$ contains the increasing exponent
${\rm e}^{2\epsilon^2 l}$ and oscillating terms, decaying
as ${\rm e}^{-\epsilon^2 l}$. The analogous picture is valid
for higher moments: the maximal exponent $x_n^{max}$ in the
metallic regime is real (see (43)) and
is not assisted by
oscillations, while the oscillating terms grow more slowly and
are relatively small even for positive $\alpha_s$. As a result,
the
situation for a typical value of $\rho$ is qualitatively the
same as for its moments and, coming to dimensionless
conductance $g=1/\rho$, one finds its oscillations being
decreasing. If the typical values $\Delta_1\sim \Delta_2\sim 1$
are accepted for the foreign leads, the oscillations of $g$ are
of the order of unity in the range $\epsilon^2 l \alt 1$, while
the dimensional conductance fluctuates on the level
of $e^2/h$ (Fig.1).  According to this picture, only the
order of magnitude of these fluctuations is universal, while
their amplitude may be strongly affected by the change of the
Fermi level and the properties of the ideal leads.

\begin{center}
{\bf 11. Conclusion}
\end{center}

The paper presents the accurate Fourier analysis of
aperiodic conductance oscillations discovered in the
classical experiments by Webb and Washburn \cite{5}. The
obtained results reconcile two alternative scenarios
discussed in Sec.\,1. On one hand, the Fourier spectrum is
practically discrete, confirming the conception of the paper
\cite{6}, according to which aperiodic oscillations are
determined by a superposition of incommensurate harmonics.
On the other hand, the spectrum in whole resembles the discrete
white noise, whose properties are close to the continuous one.
The more detailed analysis reveals  existence of the
continuous component, whose smallness is theoretically
explained in Sec.\,5.

The paper discovers a lot of qualitative moments, confirming the
presented picture. The frequencies of the discrete harmonics
depend very slightly on the treatment procedure, which proves
their objective existence. The "natural" origin of the $f(x)$
argument, established after exclusion of the shift oscillations
confirms validity of the analogy with 1D systems. The same
is confirmed by manifestations of the exponential growth
of harmonics. The phase distribution for the
coefficients $A_s$ agree with their expected randomness.
The distribution of the growth exponents and the differences of
neighboring frequencies is in agreement with  theoretical
results for the metallic regime. Microscopical estimates are in
agreement with the geometrical dimensions of the sample.

Universal conductance fluctuations are discussed in a lot of
works (see [20--40] and references therein), and it
would be interesting to process another experimental
data in the spirit of the present paper.

The author is grateful to V.V.Brazhkin, who was an
initiator of the paper \cite{11a}.

  \vspace{6mm}

\begin{center}
{\it Appendix 1.} {\bf Solution of equation (19)}
\end{center}

Let restrict the analysis by the metallic regime,
where typical values of $\rho$ are small. Consider
the eigenvalue problem for the operator in the right
hand side of Eq.\,19 and, retaining the terms of the lower
order in $\rho$,  come to equation
$$
-\lambda P = P'_\rho + \rho P''_{\rho \rho} \,.
\eqno(A.1)
$$
Assuming that $\rho$ changes in the interval from zero till
$R$, let accept the condition of finiteness at $\rho=0$ and
the zero boundary condition at $\rho=R$. Then eigenvalues and
eigenfunctions have the form
$$
\lambda_s =\mu_s^2/4R\,,\qquad
e_s(\rho)= J_0\left( \mu_s \sqrt{\rho/R} \right)\,,
\eqno(A.2)
$$
where $\mu_s$ are the roots of the Bessel function $J_0(x)$.
Solving Eq.\,19 with the initial condition (22) by expansion
of $P(\rho)$ over eigenfunctions $(A.2)$, we obtain
$$
P(\rho,t) = \int\limits_0^\infty 2\mu d\mu\, {\rm
e}^{-\mu^2 t}\, J_0(2\mu\sqrt{\rho_0})\,
J_0(2\mu\sqrt{\rho}) \,.
\eqno(A.3)
$$
Here we set $t=\alpha(L-L_0)$ and take into account that for
large $R$ the spectrum of $\mu_s$ becomes
quasi-continuous and summation over $s$ may be replaced by
integration over $\mu$,  using the asymptotics
$\mu_s=\pi s + const$ for large $s$.  Calculating the integral
in $(A.3)$, we have
$$
P(\rho,t) = \frac{1}{t}
\exp{\left\{ -\frac{\rho+\rho_0}{t} \right\}} \,
I_0\left( \frac{2\sqrt{\rho \rho_0}}{t}\right)\,,
\eqno(A.4)
$$
where $I_0(x)=J_0(ix)$. For $\rho\alt t$ and $t\gg \rho_0$
distribution $(A.4)$ transforms to (21), while for $\rho$
close to $\rho_0$ and $t\ll \rho_0$  accepts the
Gaussian form
$$
P(\rho,t) = \left( \frac{1}{4\pi \rho_0 t} \right)^{1/2}
\exp{\left\{ -\frac{ (\rho-\rho_0)^2}{4\rho_0 t} \right\}}\,.
\eqno(A.5)
$$
The closeness of $(A.4)$ to the Gaussian distribution
allows to characterize it by two first moments. Multiplying
Eq.\,19 by $\rho^n$ and integrating over $\rho$, we have the
evolution equation for the moments of the distribution
$P(\rho)$; their solution for the initial condition (22)
has a form
$$
\left\langle \rho\right\rangle = -\frac{1}{2}
+\frac{1+2\rho_0}{2}\, {\rm e}^{2t} \,,
\eqno(A.6)
$$
$$
\left\langle \rho^2\right\rangle = \frac{1}{3}
-\frac{1+2\rho_0}{2}\, {\rm e}^{2t} +
\left[\rho_0^2 +\frac{1+2\rho_0}{2} -\frac{1}{3} \right] {\rm
e}^{6t}  \,.
$$
and simplifies for small $\rho$,
$$
\left\langle \rho\right\rangle = \rho_0 + t\,,
$$
$$
\left\langle \rho^2\right\rangle = \rho^2_0 + 4\rho_0 t+2 t^2\,,
$$
$$
\sigma^2 =2 \rho_0 t + t^2\,,
\eqno(A.7)
$$
in agreement with distribution $(A.4)$.
These results are valid for $L>L_0$. Analysis of
the interval $0<L<L_0$ is complicated by necessity to satisfy
two conditions (20) and (22), which is possible only under
certain restrictions on the realization of the random potential.
Such restrictions should be imposed on the interval
$(0,L_0)$ in whole,  and are not essential for small $L$, and $L$
close to $L_0$.  In the former case we have $\langle
\rho\rangle=\sigma$ in correspondence with distribution (21). In
the latter case the situation is determined by the fact that
spreading of the distribution occurs
symmetrically\,\footnote{\,Equation (19) has the same form, if we
set $L=L_0+l$  or $L=L_0-l$ and consider evolution over $l$. It
is clear from the derivation scheme of such equations (see
Appendix $A$ in \cite{38}).} for deviations of $L$ to left or to
right from $L_0$. As for the quantity $\langle \rho\rangle$,
conditions (20) and (22) are satisfied automatically, if $\rho_0$
coincides with the mean value of distribution (21) for
$L=L_0$.  The situation does not change qualitatively, if
a typical value from distribution (21) is chosen for $\rho_0$.
As a result, we comes to the picture presented in
Fig.\,8.

  \vspace{6mm}

\begin{center}
{\it Appendix 2.} {\bf Evolution of $\langle
\rho\rangle$ for the general initial condition} \end{center}

Calculation of $\langle \rho\rangle$ is reduced technically to
the study of evolution of the second moments for the
transfer matrix with complex elements $T_{ij}$ \cite{6}
$$
z_{1}^{(l)}=\left\langle
\left| T_{11}^{(l)}\right|^2\right\rangle\,, \qquad
z_{2}^{(l)}=\left\langle
T_{11}^{(l)}  T_{12}^{(l)*}\right\rangle\,, \qquad
$$
$$
z_{3}^{(l)}=\left\langle
T_{11}^{(l) *}  T_{12}^{(l)}\right\rangle\,, \qquad
z_{4}^{(l)}=\left\langle
\left|T_{12}^{(l)}\right|^2\right\rangle\,, \qquad
 \eqno(A.8)
$$
They satisfy the system of difference equations, whose general
solution has a form \cite{6}
$$
\left ( \begin{array}{cccc} z_{1}^{(l)} \\ z_{2}^{(l)}
\\ z_{3}^{(l)} \\ z_{4}^{(l)}\end{array} \right)\,
=C_0\,\left ( \begin{array}{cccc} -1\\ 0\\ 0 \\ 1
\end{array} \right)\,
+\sum\limits_{i=1}^3\, C_i  \,
\left ( \begin{array}{cccc} 1 \\ e_{2}(x_i)
\\ e_{3}(x_i)\\ 1 \end{array} \right)\, \exp{(x_i l)} \,,
\eqno(A.9)
$$
where $x_1,\,\,x_2,\,\,x_3$ are the roots of the first
equation (37), and
$$
e_{2}(x)=\frac{{\cal A}x+{\cal B}}{p(x)}\,,\qquad
e_{3}(x)=\frac{{\cal A}^* x+{\cal B}^*}{p(x)}\,,
$$
$$
{\cal A}=2\epsilon^2 -2i\Delta\,,
\qquad {\cal B}=4\alpha\Delta + 4i\epsilon^2 (\alpha-\Delta)\,,
\eqno(A.10)
$$
$$
p(x)=x^2+2\epsilon^2 x +4 \alpha^2 \,.
$$
Here $\alpha=-\Delta_2 \delta$, $\Delta=\Delta_1 \delta$,
while $\delta$ and $\epsilon^2$ are defined after (23), and
$$
\Delta_1=\frac{1}{2}
\left(\frac{k}{\bar k}-\frac{\bar k}{k} \right)\,,\qquad
\Delta_2=\frac{1}{2}
\left(\frac{k}{\bar k}+\frac{\bar k}{k} \right)\,.
\eqno(A.11)
$$
where $\bar k$ and $k$ are Fermi momenta in the system under
consideration and in the ideal leads connected to it. In
contrast to \cite{6}, we accept the initial condition
not in the form of the unit matrix, but as the general transfer
matrix
%
%
$$
 T=  \left ( \begin{array}{cc} \sqrt{\rho\!+\!1}\, e^{i\varphi} &
\sqrt{\rho} \,e^{i\theta}
\\ \sqrt{\rho}\, e^{-i\theta} & \sqrt{\rho\!+\!1}\, e^{-i\varphi}
\end{array} \right)\,,
\eqno(A.12)
$$
with $\rho=\rho_0$ and $\psi=\theta-\varphi$. The quantity
$z_4^{(l)}$ immediately determines $\langle \rho\rangle$,
and the general result for the latter has a form
$$
\langle \rho\rangle = C_0 + C_1 \, {\rm e}^{x_1 l}
+ C_2 \, {\rm e}^{x_2 l}+ C_3 \, {\rm e}^{x_3 l} \,,
\eqno(A.13)
$$
where
$$
C_0=-\frac{1}{2}\,,\qquad
C_i=(-1)^{i+1} \left[ \,
(1\!+\!2\rho_0)\,\frac{Q_i}{2Q}\,+ \right.
$$
$$
\left.
+K_1\,\frac{R_i}{Q}\,+K_2\,\frac{S_i}{Q}\,
 \right]\,,
\quad  i=1,2,3
\eqno(A.14)
$$
and
$$
Q_1=(x_2\!-\!x_3)\,p(x_1)\,,\quad
Q_2=(x_1\!-\!x_3)\,p(x_2)\,,
$$
$$
Q_3=(x_1\!-\!x_2)\,p(x_3)\,,
$$
$$
Q = Q_1- Q_2+Q_3=
x_1^2\, (x_2\!-\!x_3) - x_2^2\, (x_1\!-\!x_3) +
x_3^2\, (x_1\!-\!x_2) \,.
$$
$$
R_1=\left[x_2p(x_3)\!-\!x_3p(x_2)\right]\,p(x_1)\,,
$$
$$
R_2=\left[x_1p(x_3)\!-\!x_3p(x_1)\right]\,p(x_2)\,,
$$
$$
R_3=\left[x_1p(x_2)\!-\!x_2p(x_1)\right]\,p(x_3)\,,
$$
$$
S_1=\left[p(x_3)\!-\!p(x_2)\right]\,p(x_1)\,,
$$
$$
S_2=\left[p(x_3)\!-\!p(x_1)\right]\,p(x_2)\,,
$$
$$
S_3=\left[p(x_2)\!-\!p(x_1)\right]\,p(x_3)\,,
$$
$$ K_1= \frac{\epsilon^2\sin{\psi}-\Delta\cos{\psi}}
 { 4\left[ \alpha\Delta^2+\epsilon^4(\alpha-\Delta)
 \right] }\, \sqrt{\rho_0(1+\rho_0)}  \,,
\eqno(A.15)
 $$
$$ K_2=  \frac{\epsilon^2(\alpha-\Delta)\cos{\psi}
 +\alpha\Delta\sin{\psi}}
 { 2\left[ \alpha\Delta^2+\epsilon^4(\alpha-\Delta)
 \right] } \,\sqrt{\rho_0(1+\rho_0)}  \,.
 $$
Using asymptotic forms for $x_1,\,\,x_2,\,\,x_3$ in the metallic
regime \cite{6},  we come to results (23) and
(25). The first result is valid for the "natural" ideal
leads\,\footnote{\,In this case, the roots $x_1,\,\,x_2,\,\,x_3$
should be expanded in $\epsilon^2/\delta$ to the higher
order than in the paper \cite{6}.},
which differ from the system under consideration only by
absence of the random potential in them; in this case
$k=\bar k$, $\Delta_1=0$ and oscillations occur in the first
order in the small parameter $\epsilon^2/\delta$. Result
(25) is valid for foreign leads, when $\Delta_1\ne 0$ and
oscillations occur in the zero order in $\epsilon^2/\delta$.
The use of the asymptotic results for $x_1,\,\,x_2,\,\,x_3$
in the  "critical" region \cite{6} leads to result (24),
applicable near the edge of the initial band; it is given for
the "natural" leads, since the situation for the foreign
leads is sufficiently illustrated by the formulas presented
in \cite{6}.

\begin{center}
{\it Appendix 3.} {\bf On the choice of the natural $x$
origin} \end{center}

According to the Onsager relations, conductance is an even
function of the magnetic fiel $B$ and accepting $x=B$
one has an even function $f(x)$ in Eq.3. Let accept the
smoothing function in the $x$-symmetrized form
$G(x\!-\!a)+G(x\!+\!a)$ with even $G(x)$.  Then the Fourier
transform is real and apart the sign coincides with its modulus,
so it does not contain the shift oscillations. Now let remove the
function $G(x\!+\!a)$. The arising Fourier transform $F(\omega)$
of the function  $f(x) G(x\!-\!a)$ appears to be complex, with
its real part being half of the previous\,\footnote{\,One can
verify easily that the Fourier transforms of functions $f(x)
G(x\!+\!a)$ and $f(x) G(x\!-\!a)$ have equal real and opposite
imaginary parts.}, and free of the shift oscillations.
The latter will be absent also in ${\rm Im} \, F(\omega)$, since
they affect equally the real and imaginary parts. After the
shift $x\to x\!+\!a$ the Fourier transforn looks as follows
$$
F(\omega)={\rm e}^{i\omega a} \int f(x\!+\!a) G(x)
{\rm e}^{i\omega x} dx
$$
and the arising integral for $a=\mu_0$ corresponds to the
integral considered in Sec.2, while the factor ${\rm
e}^{i\omega a}$ produces shift oscillations. However,
the obtained sign of $a$ is opposite to that found
empirically.

The origin of the contradiction lies in the fact that the
Onsager symmetry distinguishes not only value $B=0$ but also
$B=\infty$, and just the latter corresponds to the empirical
situation. Indeed, accepting $x=1/B$, we can repeat the
previous argumentation, but now the decrements of $B$
and $x$ have opposite signs, and for the qualitative
correspondence with Sec.2 the sign of $\omega$ should be changed.
As a result, the factor $\exp\{i\omega a\}$ has a correct
sign of $a$.


\end{document}